\documentclass[preprint,5p,twocolumn]{elsarticle}
\usepackage{amssymb}
\usepackage{amsmath}
\usepackage{mathrsfs}
\usepackage{natbib}

\usepackage{epsfig}
\usepackage{graphics}
\usepackage{float}
\usepackage{tabularx}
\usepackage{color}

\expandafter\ifx\csname package@font\endcsname\relax\else
 \expandafter\expandafter
 \expandafter\usepackage
 \expandafter\expandafter
 \expandafter{\csname package@font\endcsname}%
\fi
\newcolumntype{C}[1]{>{\centering\arraybackslash}p{#1}}

\begin{document}
 
\title{Plateau Inflation in SUGRA-MSSM}
\author[PRL]{Girish Kumar Chakravarty}
\ead{girish20@prl.res.in}
\author[PRL]{Gaveshna Gupta}
\ead{gaveshna@prl.res.in}
\author[DFCU,INFN]{Gaetano Lambiase}
\ead{lambiase@sa.infn.it}
\author[PRL]{Subhendra Mohanty}
\ead{mohanty@prl.res.in}

\address[PRL]{Theoretical Physics Division, Physical Research Laboratory, Ahmedabad 380009, India.}
\address[DFCU]{Dipartimento di Fisica "E.R. Caianiello" Universit\'a di Salerno, I-84084 Fisciano (Sa), Italy.}
\address[INFN]{INFN - Gruppo Collegato di Salerno, Italy.}

\def\be{\begin{equation}}
\def\ee{\end{equation}}
\def\al{\alpha}
\def\ba{\begin{eqnarray}}
\def\ea{\end{eqnarray}}
\def\beas{\begin{eqnarray*}}
\def\eeas{\end{eqnarray*}}


\begin{abstract}
We explored a Higgs inflationary scenario in the SUGRA embedding of the MSSM in Einstein frame where the inflaton
is contained in the $SU(2)$ Higgs doublet. We include all higher order non-renormalizable terms to the MSSM 
superpotential and an appropriate K\"ahler potential which can provide slow-roll inflaton potential in the
$D-$flat direction. In this model, a plateau-like inflation potential can be obtained if the imaginary part
of the neutral Higgs acts as the inflaton. The inflationary predictions of this model are consistent with the
latest CMB observations. The model represents a successful Higgs inflation scenario in the context of Supergravity
and it is compatible with Minimal Supersymmetric extension of the Standard Model.

\end{abstract}


\maketitle

\section{Introduction}
Observations of super-horizon  ansiotropies in the CMB (COBE, WMAP, Planck) have established that
early universe underwent a period of cosmic inflation \cite{inflation}. Such a period of rapid expansion can
solve a number of cosmological problems, such as the horizon, flatness and monopole problems, and generate the initial 
conditions (homogeneity and isotropy) for the hot big bang evolution of the universe thereafter. It not only
explains the scale-invariant and Gaussian spectrum of density fluctuations on superhorizon scales but also 
provides the seed for the large-scale structures formation in the universe. In particle physics, 
the issues like dark matter, hierarchy problem, baryogenesis and non-renormalizability of gravity etc. 
perpetuates and hints towards the existence of new physics beyond standard model. Till date the most promising approach 
to address these key issues is local supersymmetry also known as supergravity (SUGRA). 

In the framework of global supersymmetry (SUSY), there have been attempts to construct Higgs
field driven inflation models in MSSM (minimal supersymmetric standard model) and NMSSM 
(next-to-minimal supersymmetric standard model) with or without non-minimal coupling to curvature
~\cite{Chatterjee:2011qr,Ibanez:2014kia,Ibanez:2014swa,Nakayama:2010sk,Allahverdi:2006cx,
Allahverdi:2007wt,Mazumdar:2010sa}.
Within the SUGRA framework, the non-minimal and minimal Higgs inflation model in MSSM and NMSSM are studied in
~\cite{Einhorn:2009bh,Ferrara:2010yw,Lee:2010hj,Terada:2015cna}. Apart from these SUGRA-(N)MSSM models, there exist
a number of inflation models in the framework of Einstein gravity and modified gravity \cite{Martin:2013tda,DeFelice:2010aj}. 
In the standard slow-roll inflationary scenario if one tries
to couple the standard model (SM) with Einstein gravity, one finds that the SM Higgs can not be identified 
as the inflaton because it has a very small self-coupling $~\mathcal{O}(10^{-13})$ and light mass 
$~\mathcal{O}(10^{13}) GeV$ at Planck scale. Apart from this it predicts large amplitude of the gravity
waves which is ruled out by the joint analysis of BICEP2/Keck Arrey and the Planck observations at $95\%CL$
~\cite{bicep2keck2015}. However, if a non-minimal interaction of the type $\xi H^{\dagger}H R$ between 
the inflaton and gravity is considered then for large value of the non-minimal
coupling parameter $\xi\sim10^{4}$, the inflaton can be identified with the SM Higgs~\cite{Fakir:1990eg,Bezrukov}.
The large $\xi$ allows the self coupling to be $~\mathcal{O}(1)$ and therefore Higgs mass $\sim 125 GeV$ at Electroweak
scale consistent with the LHC experiment~\cite{Chatrchyan:2012xdj,Aad:2012tfa}. Also this scenario predicts small gravity wave amplitude
consistent with the observations. However, in this setting higgs-graviton scattering suggest the cut off in 
the theory to be $\Lambda=Mp/\xi$ which is much smaller than the energy scale during inflation $\Lambda=Mp/\sqrt{\xi}$
due to large non-minimal coupling, and therefore this scenario suffers from unitarity violation problem. Various ways
to solve this issue are proposed in~\cite{Burgess:2009ea,Barbon:2009ya,Bezrukov:2009db,Barvinsky:2009fy}. However
at present, this model (and the equivalent Starobinsky model of inflation) is one of the most favored models of inflation 
due to its small but observationally consistent prediction of tensor to scalar ratio $r\simeq0.003$.

In the context of Jordan Frame SUGRA embedding of MSSM in presence of non-minimal interaction of Higgs, Einhorn and Jones
demonstrated that, under certain assumptions, the slow-roll conditions are not 
met along $\beta$-direction ($\beta$ being the ratio of two Higgs vevs) for non-zero $D-$term and therefore slow-roll
inflation can not be achieved. On the other hand, in $D-$flat direction the inflaton potential is negative and therefore
unsuitable for inflation. However they found that slow-roll inflation can be realized in NMSSM in which a gauge 
singlet $S$ is added to existing two Higgs doublets $(H_1, H_2)$~\cite{Einhorn:2009bh}. $\mathcal{N}=1$, $D=4$ 
Jordan frame supergravity in a superconformal approach~\cite{Kallosh:2000ve} with arbitrary scalar-curvature 
coupling is formulated in~\cite{Ferrara:2010yw,Ferrara:2010in}. The Einhorn and Jones NMSSM inflationary scenario
appears as a special case of this formulation where they showed that a strong tachyonic 
instability $(m_{\tilde{s}}^{2}>H^{2})$ in the $S-$direction during inflation because the scalar potential in 
$D-$flat direction has a saddle point at $S=0$ and therefore inflaton has an unstable trajectory at $S=0$
~\cite{Ferrara:2010yw}. Later on it was shown in \cite{Lee:2010hj}, that a higher order correction of the type
$-\gamma (S^{\dagger}S)^{2}$ to the frame function can cure the problem of tachyonic instability if, for 
$\gamma\gtrsim0.003$, one chooses a very small cubic coupling $\sim\mathcal{O}(10^{-5})$ of the gauge singlet 
in the superpotential. Also the unitarity problem, which exists even in Supergravity generalisation
~\cite{Einhorn:2009bh} of standard non-minimal Higgs inflation scenario\cite{Bezrukov}, seems to be resolved here.
The possibility of Higgs inflation in MSSM in context of supergravity with large Higgs field and fractional
power potential has been explored in~\cite{Terada:2015cna}.

In the present work we study an inflation model in the SUGRA embedding of the MSSM in Einstein frame.
As this will be a minimal SUGRA-MSSM model, so there will not arise issues like tachyonic instabilty and unitarity violation 
during slow-roll inflationary regime. Unlike the Einhorn and Jones MSSM-SUGRA inflationary scenario, in this
model, a $D-$flat positive inflaton potential can be achieved by adding higher order non-renormalizable terms 
to the MSSM superpotential $\mu H_u \cdot H_d$. And to obtain the correct inflationary observables, 
the required flatness of the inflation potential can be achieved when the imaginary part of the neutral 
Higgs component in the Einstein frame acts as the inflaton.

The organisation of the remainder of the paper is as follows. In Section~\S\ref{model}, we introduce the model and calculate
the $F-$term and $D-$term inflaton potential. Then to constrain the inflationary observables, we derive the effective
potential in canonical inflaton field basis. In Section~\S\ref{inflation_parameters}, we present the model predictions of 
inflationary observables : in particular spectral index $n_s$ and its
running $\alpha_s$, tensor-to-scalar ratio $r$ and constrain the couplings in the model from CMB normalisation. Also we discuss
the possibility of the slow-roll potential with respect to the field $\beta$.
Finally, we present our conclusions in Section~\S\ref{conclusion}.


\section{The Model}\label{model}
In this model we consider the following K\"ahler potential $K(\phi_{i},\phi^{*}_{i})$  
\ba
 K &=& 3 M_{p}^{2} \ln \left[1 + \frac{1}{3 M_{p}^{2}}\left(H^{\dagger}_{u} H_{u} + H^{\dagger}_{d} H_{d}\right)\right],\label{KP1} 
 \ea
and superpotential $W(\phi_{i})$ with higher order non-renormalizable terms
\be
W = \mu(H_{u}\cdot H_{d})+\lambda \frac{(H_{u}\cdot H_{d})^{2}}{M_{p}} \exp\bigg(\frac{H_{u}\cdot H_{d}}{M_{p}^{2}}\bigg) \,,\label{SP1}
\ee
where $H_u$ and $H_d$ are $SU(2)$ Higgs doublets identified as up-type and down-type
Higgs superfields, given by
\be
 H_{u} =\begin{pmatrix}
  \phi_{u}^{+}  \\
  \phi_{u}^{0}
 \end{pmatrix}\,,~~~~
 H_{d} =\begin{pmatrix}
  \phi_{d}^{0}  \\
  \phi_{d}^{-}
 \end{pmatrix}\,,\label{HuHd}
\ee
and the contraction $H_{u}\cdot H_{d}$ is the $SU(2)$ invariant $H_{u}\cdot 
H_{d}\equiv H_{u}^{T} i\sigma_2 H_{d}=\phi_{u}^{+} \phi_{d}^{-}-\phi_{u}^{0} \phi_{d}^{0}$.
Considering only the neutral components of $H_{u}$ and $H_{d}$ to be non-vanishing, we obtain
$H_{u}\cdot H_{d}=-\phi_{u}^{0} \phi_{d}^{0}$. \footnote{For simplicity, we shall omit the superscript `0'
and work in $M_{p}=(8\pi G)^{-1/2}=1$ unit from here onwards.} The first term $\mu(H_{u}\cdot H_{d})$
in~(\ref{SP1}) is the MSSM superpotential contains a parameter $\mu$ of the order of electroweak 
scale $\sim O(100)~GeV$ whereas Higgs fields are of the order of Planck scale during inflation.
Therefore we will neglect the first term in $W$ compared to the second term which includes all higher order 
terms in $H_{u}\cdot H_{d}$. 

The scalar potential in SUGRA depends upon the 
K\"ahler function $G(\phi_{i},\phi^{*}_{i})$ given in terms of superpotential 
$W(\phi_{i})$ and K\"ahler potential $K(\phi_{i},\phi^{*}_{i})$ as
\be
G(\phi_{i},\phi^{*}_{i}) \equiv K(\phi_{i},\phi^{*}_{i}) + \ln W(\phi_{i}) +\ln W^{\ast}(\phi^{*}_{i}),
\ee
where $\phi_{i}$ are the chiral scalar superfields. 
The scalar potential in Einstein frame is given as $V=V_F + V_D$, where the F-term potential is given by
\be
V_{F}=e^{G}\left[\frac{\partial G}{\partial \phi^{i}} K^{i}_{j*} \frac{\partial G}{\partial \phi^{*}_{j}} - 3 \right] \label{LV}
\ee
and the D-term potential is given by
\be
V_D= \frac{1}{2}\left[Re f_{ab}^{-1}(\phi_{i})\right] D^{a}D^{b},\label{VD}
\ee
where 
\be
D^{a}=-g \frac{\partial G}{\partial \phi_{k}}(\tau^{a})^{l}_{k}\phi_{l}
\ee
and $f_{ab}$ is related to the kinetic energy of the gauge field thus it must
be a holomorphic function of $\phi_i$. $g$ is the gauge coupling
constant corresponding to each gauge group and $\tau^{a}$ being the corresponding generator. 
For $SU(2)_L$ symmetry
$\tau^{a}=\sigma^{a}/2,$
 where $\sigma^{a}$ are Pauli matrices and for $U(1)_{Y}$ symmetry, $\tau^{a}$ are hypercharge of the fields, 
 i.e. $Y_u= 1$ and $Y_d=-1$.

The kinetic term of the scalar fields is given by
\be
\mathcal{L}_{KE}=K_{i}^{j*} \partial_{\mu}\phi^{i} \partial^{\mu}\phi^{*}_{j} \,,\label{LK}
\ee
here $K^{i}_{j*}$ is the inverse of the K\"ahler metric
\be
 K_{i}^{j*} \equiv \frac{\partial^{2}K }{ \partial\phi^{i}\partial\phi^{*}_{j}}\,.
\ee 
 
%
Using (\ref{HuHd}),
the K\"ahler potential (\ref{KP1}) and superpotential (\ref{SP1}) 
reduce to
\be
 K = 3 \ln \left[1+\frac{\phi_{u}\phi_{u}^{*}+\phi_{d}\phi_{d}^{*}}{3}\right]\,,\label{KP}
 \ee
\be
W= \lambda (-\phi_{u}\phi_{d})^{2} \exp(-\phi_{u}\phi_{d})\,,\label{SP}
\ee
respectively.
Considering the canonical form the gauge kinetic function $f_{ab}=\delta_{ab}$ for simplicity, the D-term potential becomes
\be
V_D= \frac{(g_{1}^{2}+g_{2}^{2})}{8\left(1+\frac{\phi_{u}^{\ast}\phi_{u}+\phi_{d}^{\ast}\phi_{d}}{3}\right)^{2}}
\left[\phi_{u}^{\ast}\phi_{u}-\phi_{d}^{\ast}\phi_{d}\right]^{2},
\ee
where $g_{1}$ and $g_{2}$ are gauge couplings of $U(1)_Y$ and $SU(2)_L$ symmetries, respectively.
It is convenient to parametrize the complex fields $\phi_u$ and $\phi_d$ as $\phi_{u}=\phi \sin(\beta)$ and 
$\phi_{d}=\phi \cos(\beta)$. Here, we shall treat $\phi$ as a complex field and $\beta$ to be real field.
For the given parametrization, the $D$-term potential can be given as 
\be
V_{D}=\frac{9}{8}(g_1^2 + g_2^2) \left|\phi\right|^4 \frac{\cos(2\beta)^2}
 {(3 + \left|\phi\right|^2)^2}\label{VDbeta}
\ee
and $F$-term potential can be given as
\ba
\hspace{-0.9cm}V_{F} &=& \frac{\lambda^{2}}{31104} \sin(2\beta)^2 \exp\left(-(\phi^2+\phi^{*2})\sin(2\beta)/2\right) \nonumber\\
&& \left|\phi\right|^6 (3 + \left|\phi\right|^2)^3 \bigg[1152 + \left|\phi\right|^2 \Big\{756 + \left|\phi\right|^2 \big[232 \nonumber\\
&&   + 3\left|\phi\right|^2 (7 + 9 \left|\phi\right|^2)\big]\Big\}- 4 \left|\phi\right|^2 \Big\{93 + \left|\phi\right|^2 \big[58  \nonumber\\
&&  + \left|\phi\right|^2(6+\left|\phi\right|^2)\big]\Big\} \cos(4\beta)+(3+\left|\phi\right|^2) \Big\{ \left|\phi\right|^6 \nonumber\\
&&  \cos(8\beta)-8(\phi^2+\phi^{*2})\big[12 +7 \left|\phi\right|^2 \sin(2\beta)^{2} \big]\Big\} \nonumber\\
&&   \sin(2\beta)\bigg]\,. \label{Lvf}
\ea

Now we calculate the kinetic term (\ref{LK}) which comes out to be
\be
\mathcal{L}_{KE}= 
\frac{9}{(3+|\phi|^{2})^{2}}|\partial_{\mu}\phi|^{2}\,,\label{Lke}
\ee
 
In order to make the kinetic term ${\cal{L}}_{KE}$ canonical, we redefine the field $\phi$ to $\chi_{E}$ via
\be
\left|\partial_{\mu}\chi_{E} \right|^{2} = \frac{9}{(3+\left|\phi\right|^{2})^{2}}\left|\partial_{\mu}\phi 
\right|^{2}\,. \label{chitophi}
\ee
It is straightforward to solve (\ref{chitophi}) to get
\be
\phi=\sqrt{3} \tan\left(\frac{\chi_{E}}{\sqrt{3}}\right)\,. \label{phisol}
\ee

We further decompose $\phi$ in terms of its real and imaginary parts as
$\phi=(\phi_{R}+i\phi_{I})/\sqrt{2}$ and assume its real part to be 
zero, we have $\phi=-\phi^{\ast}=i \phi_{I}/\sqrt{2}$.
Similarly, $\chi_{E}=(\chi_{R}+i\chi)/\sqrt{2}$ and setting the real part to zero, we have
$\chi_{E}=-\chi_{E}^{\ast}=i \chi/\sqrt{2}$. The solution (\ref{phisol}) can be rewritten as
\be
\phi_{I}=\sqrt{6} \tanh\left(\frac{\chi}{\sqrt{6}}\right)\,.\label{phichi}
\ee
where we have used the trigonometric relation $\tan(i \theta)=i \tanh(\theta)$. Here $\chi=\sqrt{2}Im(\chi_E)$ acts as the inflaton.
If we assume the real part of the fields to be non-zero and imaginary parts to be zero, the potential
becomes very steep due to $\phi\propto \tan(\chi_R)$ during inflationary regime, therefore unsuitable for 
slow-roll inflation. However, we will see that the imaginary part~(\ref{phichi}) can provide
the required slow-roll inflaton potential.

From the analysis of the potential we find  that the slow-roll conditions are met only when the D-term 
potential vanishes which is achieved by choosing $tan(\beta)=1$. In terms of canonical field $\chi$, 
substituting from (\ref{phichi}) into (\ref{Lvf}), the $D$-flat inflaton
potential becomes
\ba
\hspace{-.5cm}U(\chi)&=&\frac{27\lambda^{2}}{16} \exp\left(3 \tanh\big(\chi/\sqrt{6}\big)^{2}\right) 
\tanh\big(\chi/\sqrt{6}\big)^{6}\nonumber \\ 
&&\Big[1+\tanh\big(\chi/\sqrt{6}\big)^{2}\Big]^3 \Big[16+71\tanh\big(\chi/\sqrt{6}\big)^{2} \nonumber \\
&&  +124 \tanh\big(\chi/\sqrt{6}\big)^{4} +60\tanh\big(\chi/\sqrt{6}\big)^{6} \nonumber \\
&& +9\tanh\big(\chi/\sqrt{6}\big)^{8} \Big]\,.\label{Upot}
\ea
\section{Model predictions of the inflationary observables and the dynamics of the field $\beta$}\label{inflation_parameters}
In this Section we estimate the inflationary observables for the model discussed above.
In the Einstein frame, the slow-roll parameters are defined as
\be
\epsilon \equiv \frac{1}{2}\left(\frac{U'}{U}\right)^2\,,~~~~~~~\eta \equiv \frac{U''}{U}\,,~~~~~~~\xi \equiv \frac{U'U'''}{U^{2}}\,.
\ee
Inflation ends when the condition $\epsilon(\chi_e)=1$ is met, which determines the field value at the end inflation
$\chi_e$. Using the e-folding expression 
\be
N=\int_{\chi_e}^{\chi_s} \frac{U}{U'} d\chi\,,\label{efolds}
\ee 
we can obtain the initial field value $\chi_{s}$ corresponding to $N \approx 60$ e-folds before the end of inflation,
when observable CMB modes leave the horizon.

For the estimation of the inflationary observables tensor to scalar ratio $r$, scalar spectral index $n_{s}$ and
running of spectral index $\alpha_s$, we use the standard Einstein frame relations, given by
\ba
r &=& 16\epsilon\,,\\
n_{s} &=& 1-6\epsilon+2\eta\,,\\
\alpha_{s} &\equiv& \frac{dn_{s}}{d\ln k} = 16\epsilon\eta -24\epsilon^{2} - 2\xi\,.
\ea
The coupling parameter $\lambda$ can be estimated using the CMB normalisation $i.e.$ using the
standard expression for the amplitude of the curvature perturbation, given by
\be
\Delta_{\mathcal R}^{2} = \frac{1}{24 \pi^{2}} \frac{U}{\epsilon}\,.
\ee
The Planck-2015 observations give the scalar amplitude and the scalar spectral index as 
$10^{10}\ln(\Delta_{\mathcal R}^{2}) = 3.089\pm 0.036$ and $n_{s}= 0.9666 \pm 0.0062$ respectively
at ($68 \%$ CL, PlanckTT+lowP) \cite{Planck:2015,Ade:2015lrj}. The constraint on the running of the spectral 
index is $\alpha_s = - 0.0084 \pm 0.0082$ $~(68 \% CL, Planck TT+lowP)$~\cite{Ade:2015lrj}. Also, the Planck analysis of full CMB
polarization and temperature data combined with BICEP2/Keck Array CMB polarization observations have 
put an upper bound on tensor-to-scalar ratio $r_{0.05} < 0.07 \,(95\%$ CL)~\cite{bicep2keck2015}. 

\begin{figure}[t]
\centering
\includegraphics[width=8cm]{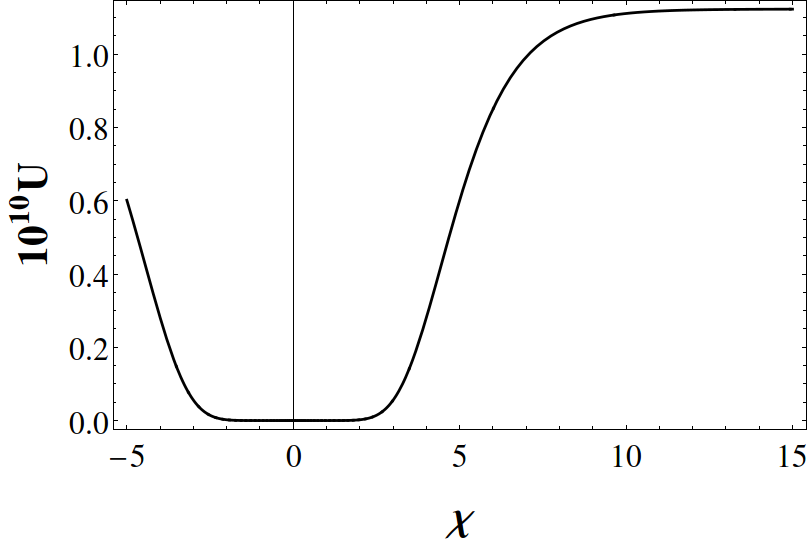}
\caption{The shape of the $D-$flat (when $\beta=\frac{\pi}{4}$) inflaton Potential is shown for $\lambda\simeq3.8\times10^{-8}$.}
 \label{fig1}
\end{figure}

\begin{figure}[t]
\includegraphics[width=9cm]{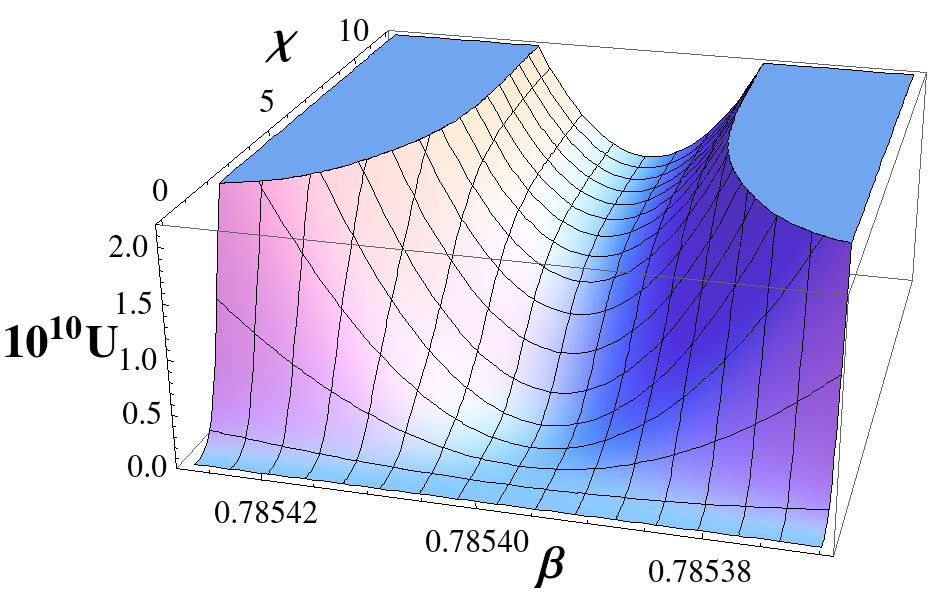}
\caption{During $60$ e-folds inflation, when the inflaton field rolls from $\chi_s\simeq8.95$ and $\chi_e\simeq3.66$,
the stabilization of the field $\beta$ at $\beta=\frac{\pi}{4}$ is shown. The slow-roll inflation
takes place along the $\chi-$direction.}
 \label{fig2}
\end{figure}

Armed with the theoretical and observational results for the CMB observables, and the inflaton potential (\ref{Upot}), 
we perform the numerical analysis of the model. For the field value $\chi_{s}\simeq8.95$ and coupling
$\lambda\simeq3.8\times10^{-8}$, we obtain $r \simeq 0.00337$, $n_{s} \simeq 0.966$ and 
$\alpha_{s} \simeq -5.7\times10^{-4}$ which are consistent with the CMB observations. And using the condition
$\epsilon(\chi_e)=1$, we obtain the field value at the end of inflation to be $\chi_e\simeq3.66$. Also from the 
e-folding expression~(\ref{efolds}), for $\chi_{s}\simeq8.95$ and $\chi_e\simeq3.66$, we obtain the minimum 
required e-folds $N\approx59$. The shape of the inflaton potential along $D-$flat direction is shown in 
Fig.~(\ref{fig1}) and the potential with varying $\beta$ and $\chi$ fields is 
shown in Fig.~(\ref{fig2}).

One could also ask the possibility of a slow-roll potential along the field $\beta-$direction for some large 
fixed $\chi$, if $\beta\neq\frac{\pi}{4}$~\cite{Einhorn:2009bh}.
We study the dynamics of the field $\beta$ around the minima $\beta=\pi/4$ of the total potential $V=V_D+V_F$, evaluated in terms
of $\chi$ using eq.(\ref{phichi}), and
find that the field $\beta$ does not satisfy the slow-roll conditions $\epsilon_\beta\ll1$ and $\eta_\beta\ll1$
altogether. The slow-roll parameters $\epsilon_\beta$ and $\eta_\beta$ are defined with respect to the field $\beta$ as
\ba
\epsilon_\beta &\equiv& \frac{1}{2}\left(\frac{V'}{V}\right)^2 \,,\\
\eta_\beta &\equiv& \frac{V''}{V}= \frac{m_{\beta}^{2}}{V}\,.
\ea
where $m_{\beta}^{2}$ is the effective mass squared of the fluctuations of the field $\beta$.
With the numerical analysis, we find that in the limit $\beta\rightarrow\frac{\pi}{4}$ although 
$\epsilon_\beta\rightarrow0$, the parameter $\eta_\beta\sim10^{10}$ because during inflation mass
of the $\beta$-field $m_{\beta}\approx1.2$ is much larger than the Hubble parameter 
$H\approx6\times10^{-6}$~\cite{Hetz:2016ics}. Therefore, there is no slow-roll along $\beta-$direction, instead the field $\beta$
rapidly rolls down to the minima of the potential at $\beta=\frac{\pi}{4}$ and stays there during inflation. The inflation
takes place along the $\chi-$direction where the slow-roll conditions $\epsilon\simeq2.1\times10^{-4}$ and $|\eta|\simeq0.016$ hold good.
At the end of inflation there is reheating and the Higgs potential assumes the finite temperature values. In the electroweak era the potential
can settle to a minima where $tan(\beta)\neq 1$.

\section{Conclusion}\label{conclusion}

In this paper we have studied a Higgs inflation model in the SUGRA embedding of the MSSM.
We include all higher order non-renormalizable terms to the MSSM superpotential.
We find that the inclusion of such higher order terms in the superpotential can provide the positive inflaton
potential in $D-$flat directions. 
 In order to obtain a plateau-like inflaton
potential, which can produce the correct inflationary observables, we have to take the real part of the canonical field $\chi_E$
to be zero and its imaginary part $\chi$ acts as the inflaton.
Slow-roll analysis of the model with small superpotential coupling $\lambda\simeq3.8\times10^{-8}$ and superplanckian field value
$\chi_{s}\simeq8.95$,
provide $N\simeq59$ e-folds and the inflationary observables $r\simeq0.00337$, $n_s \simeq 0.966$,
$\alpha_s\simeq-5.7\times10^{-4}$, $\Delta_{\mathcal R}^{2}\simeq2.19\times10^{-9}$ consistent with
the latest CMB observations. We also discussed the 
possibility of a slow-roll potential with respect to the field $\beta$ for large fixed $\chi$,
and found that the slow-roll parameter with respect to the field $\beta$ violate the condition $\eta_\beta\ll1$, 
therefore slow-roll potential along the $\beta-$direction is not possible, instead
the field $\beta$ rapidly falls towards the minima and stabilize at $\beta=\pi/4$ and stays there during the entire
period of inflation. This model represents a successful Higgs inflationary scenario in the SUGRA-MSSM theory.

\section*{Acknowledgement}
We thank the anonymous referee for a very useful suggestion which we incorporated in the paper.


\begin{thebibliography}{unsrt} 
  
\bibitem{inflation}
  A.~H.~Guth,
  Phys.\ Rev.\  {\bf D23}, 347-356 (1981);
  K.~Sato,
  Mon.\ Not.\ Roy.\ Astron.\ Soc.\  {\bf 195}, 467-479 (1981);
  A.~A.~Starobinsky,
  Phys.\ Lett.\  B {\bf 91}, 99 (1980).
  
\bibitem{Chatterjee:2011qr} 
  A.~Chatterjee and A.~Mazumdar,
  JCAP {\bf 1109}, 009 (2011)
  arXiv:1103.5758.  
  
 
\bibitem{Ibanez:2014kia} 
  L.~E.~Ibáñez and I.~Valenzuela,
  Phys.\ Lett.\ B {\bf 736}, 226 (2014)
  doi:10.1016/j.physletb.2014.07.020
  [arXiv:1404.5235 [hep-th]].
  
\bibitem{Ibanez:2014swa} 
  L.~E.~Ibanez, F.~Marchesano and I.~Valenzuela,
  JHEP {\bf 1501}, 128 (2015)
  doi:10.1007/JHEP01(2015)128
  [arXiv:1411.5380 [hep-th]].
  
    
\bibitem{Nakayama:2010sk} 
  K.~Nakayama and F.~Takahashi,
  JCAP {\bf 1102}, 010 (2011)
  doi:10.1088/1475-7516/2011/02/010
  [arXiv:1008.4457 [hep-ph]].
  
\bibitem{Allahverdi:2006cx} 
  R.~Allahverdi, A.~Kusenko and A.~Mazumdar,
  JCAP {\bf 0707}, 018 (2007)
  doi:10.1088/1475-7516/2007/07/018
  [hep-ph/0608138].

\bibitem{Allahverdi:2007wt} 
  R.~Allahverdi, B.~Dutta and A.~Mazumdar,
  Phys.\ Rev.\ Lett.\  {\bf 99}, 261301 (2007)
  doi:10.1103/PhysRevLett.99.261301
  [arXiv:0708.3983 [hep-ph]].
  
  
\bibitem{Mazumdar:2010sa} 
  A.~Mazumdar and J.~Rocher,
  Phys.\ Rept.\  {\bf 497}, 85 (2011)
  doi:10.1016/j.physrep.2010.08.001
  [arXiv:1001.0993 [hep-ph]].

 
\bibitem{Einhorn:2009bh} 
  M.~B.~Einhorn and D.~R.~T.~Jones,
  JHEP {\bf 1003}, 026 (2010)
  doi:10.1007/JHEP03(2010)026
  [arXiv:0912.2718 [hep-ph]].

   
\bibitem{Ferrara:2010yw} 
  S.~Ferrara, R.~Kallosh, A.~Linde, A.~Marrani and A.~Van Proeyen,
  Phys.\ Rev.\ D {\bf 82}, 045003 (2010)
  doi:10.1103/PhysRevD.82.045003
  [arXiv:1004.0712 [hep-th]].
  
\bibitem{Lee:2010hj} 
  H.~M.~Lee,
  JCAP {\bf 1008}, 003 (2010)
  doi:10.1088/1475-7516/2010/08/003
  [arXiv:1005.2735 [hep-ph]].
  
\bibitem{Terada:2015cna} 
  T.~Terada,
  arXiv:1504.06230 [hep-ph].
  
  
\bibitem{Martin:2013tda} 
  J.~Martin, C.~Ringeval and V.~Vennin,
  Phys.\ Dark Univ.\  {\bf 5-6}, 75 (2014)
  doi:10.1016/j.dark.2014.01.003
  [arXiv:1303.3787 [astro-ph.CO]].

\bibitem{DeFelice:2010aj} 
  A.~De Felice and S.~Tsujikawa,
  Living Rev.\ Rel.\  {\bf 13}, 3 (2010)
  doi:10.12942/lrr-2010-3
  [arXiv:1002.4928 [gr-qc]].
  
\bibitem{bicep2keck2015} 
  P.~A.~R.~Ade {\it et al.},
  arXiv:submit/1390175 [astro-ph.CO].
  
  
\bibitem{Fakir:1990eg} 
  R.~Fakir and W.~G.~Unruh,
  Phys.\ Rev.\ D {\bf 41}, 1783 (1990).
  doi:10.1103/PhysRevD.41.1783

  
  \bibitem{Bezrukov}
  F.~L.~Bezrukov and M.~Shaposhnikov,
   Phys.\ Lett.\ B {\bf 659}, 703 (2008).

\bibitem{Chatrchyan:2012xdj} 
  S.~Chatrchyan {\it et al.} [CMS Collaboration],
  Phys.\ Lett.\ B {\bf 716}, 30 (2012)
  doi:10.1016/j.physletb.2012.08.021
  [arXiv:1207.7235 [hep-ex]].
  
\bibitem{Aad:2012tfa} 
  G.~Aad {\it et al.} [ATLAS Collaboration],
  Phys.\ Lett.\ B {\bf 716}, 1 (2012)
  doi:10.1016/j.physletb.2012.08.020
  [arXiv:1207.7214 [hep-ex]].

\bibitem{Burgess:2009ea} 
  C.~P.~Burgess, H.~M.~Lee and M.~Trott,
  JHEP {\bf 0909}, 103 (2009)
  doi:10.1088/1126-6708/2009/09/103
  [arXiv:0902.4465 [hep-ph]].
  
\bibitem{Barbon:2009ya} 
  J.~L.~F.~Barbon and J.~R.~Espinosa,
  Phys.\ Rev.\ D {\bf 79}, 081302 (2009)
  doi:10.1103/PhysRevD.79.081302
  [arXiv:0903.0355 [hep-ph]].
  
\bibitem{Bezrukov:2009db} 
  F.~Bezrukov and M.~Shaposhnikov,
  JHEP {\bf 0907}, 089 (2009)
  doi:10.1088/1126-6708/2009/07/089
  [arXiv:0904.1537 [hep-ph]].

\bibitem{Barvinsky:2009fy} 
  A.~O.~Barvinsky, A.~Y.~Kamenshchik, C.~Kiefer, A.~A.~Starobinsky and C.~Steinwachs,
  JCAP {\bf 0912}, 003 (2009)
  doi:10.1088/1475-7516/2009/12/003
  [arXiv:0904.1698 [hep-ph]].
  
\bibitem{Kallosh:2000ve} 
  R.~Kallosh, L.~Kofman, A.~D.~Linde and A.~Van Proeyen,
  Class.\ Quant.\ Grav.\  {\bf 17}, 4269 (2000)
  Erratum: [Class.\ Quant.\ Grav.\  {\bf 21}, 5017 (2004)]
  doi:10.1088/0264-9381/17/20/308
  [hep-th/0006179].
  
\bibitem{Ferrara:2010in} 
  S.~Ferrara, R.~Kallosh, A.~Linde, A.~Marrani and A.~Van Proeyen,
  Phys.\ Rev.\ D {\bf 83}, 025008 (2011)
  doi:10.1103/PhysRevD.83.025008
  [arXiv:1008.2942 [hep-th]].
  
\bibitem{Planck:2015} 
  P.~A.~R.~Ade {\it et al.},
  arXiv:1502.01589 [astro-ph.CO].
  
\bibitem{Ade:2015lrj} 
  P.~A.~R.~Ade {\it et al.} [Planck Collaboration],
  arXiv:1502.02114 [astro-ph.CO].
  
\bibitem{Hetz:2016ics} 
  A.~Hetz and G.~A.~Palma,
  arXiv:1601.05457 [hep-th].
    
%
%
%

\end{thebibliography}
\end{document}